\documentclass[fleqn,10pt]{wlscirep}
\usepackage[utf8]{inputenc}
\usepackage[T1]{fontenc}
\usepackage{amsmath}
\usepackage{amssymb}
\usepackage{booktabs}
\usepackage{xcolor}
\usepackage{colortbl}
\usepackage[most]{tcolorbox}
\usepackage{graphicx}
\usepackage{float}
\usepackage{makecell}
\usepackage{multirow}
\usepackage{siunitx}
\usepackage{url}
\usepackage[colorlinks=true,allcolors=blue]{hyperref}
\graphicspath{{figures/}}
\definecolor{promptframe}{HTML}{C66A0A}
\definecolor{promptback}{HTML}{FFF8EF}
\newtcolorbox{promptbox}[1]{
  enhanced,
  breakable,
  colback=promptback,
  colframe=promptframe,
  colbacktitle=promptframe,
  coltitle=white,
  fonttitle=\bfseries\small,
  fontupper=\small,
  boxrule=0.8pt,
  arc=2pt,
  left=7pt,
  right=7pt,
  top=6pt,
  bottom=6pt,
  title={#1}
}

\title{Reward misalignment as an opportunity for materials discovery}

\author[1,+,*]{Ryotaro Okabe}
\author[2,+,*]{Shoki Kishida}
\author[3]{Abhijatmedhi Chotrattanapituk}
\author[4]{Dominic Bollettieri}
\author[5]{Takuya Isogawa}
\author[6]{Yongqiang Cheng}
\author[5,$\ddagger$,*]{Mingda Li}
\author[7,$\ddagger$,*]{Masaki Adachi}

\affil[1]{Department of Chemistry, Massachusetts Institute of Technology, Cambridge, MA 02139, USA}
\affil[2]{Harvard John A. Paulson School of Engineering and Applied Sciences, Harvard University, Cambridge, MA 02138, USA}
\affil[3]{Department of Electrical Engineering and Computer Science, Massachusetts Institute of Technology, Cambridge, MA 02139, USA}
\affil[4]{Department of Physics, Massachusetts Institute of Technology, Cambridge, MA 02139, USA}
\affil[5]{Department of Nuclear Science and Engineering, Massachusetts Institute of Technology, Cambridge, MA 02139, USA}
\affil[6]{Neutron Scattering Division, Oak Ridge National Laboratory, Oak Ridge, TN 37831, USA}
\affil[7]{Lattice Lab, Toyota Motor Corporation, Tokyo, 103-0022, Japan}
\affil[+]{equal contribution.}
\affil[$\ddagger$]{equal contribution.}
\affil[*]{Corresponding authors: rokabe@mit.edu, skishida@g.harvard.edu, mingda@mit.edu, masaki\_adachi@mail.toyota.co.jp}

\begin{abstract}
Reward misalignment is usually regarded as a failure of artificial intelligence. In materials discovery with generative AI models, however, disagreement between predefined rewards and expert judgment may reveal materials that the rewards overlook. We investigate this possibility using electronic flat bands as a prime example. Experts mostly agree with a numerical bandwidth score, but their disagreements are systematic and cannot be explained by an estimated 5\% rate of mistaken labels. We learn the experts' preferences from pairwise comparisons and use them to guide a crystal diffusion model. The learned preferences and numerical score agree on the flattest bands but diverge elsewhere. For example, the preferences favor KCo$_2$As$_2$ and CoGa, reported compounds with partially flat bands that the score would discard. The framework also extends to synthesizability and phonon properties. Our findings suggest that reward misalignment can expose opportunities for materials discovery beyond what fixed rewards are designed to find.
\end{abstract}

\usepackage{pdfpages}
\begin{document}

\flushbottom
\maketitle
\thispagestyle{empty}

\newcommand{\methodname}{H-DDO}
\newcommand{\toolname}{CrysTacit}
\section*{Introduction}

Artificial intelligence is taking on a growing share of materials discovery~\cite{cheng2026ai, merchant2023scaling, okabe2024large}. Generative models for crystals propose compounds that have not been seen before~\cite{xie2022cdvae, jiao2023diffcsp, Zeni2025MatterGen, park2025exploration, ito2026lapidaryengine}. They can be steered toward a prescribed symmetry or structural motif~\cite{okabe2026structural, jiao2024space} or toward a computed property~\cite{chen2025accelerating, adachi2026regret}, and such steering has already produced candidate superconductors~\cite{prakash2026guided, li2026agentic}. Autonomous laboratories and AI agents now carry out parts of the research loop itself~\cite{szymanski2023autonomous, lu2024aiscientist}. 
Yet, despite these advances, AI-driven discovery remains fundamentally constrained by how its objectives are defined. When materials are generated using predefined numerical rewards, the search inevitably favors what those rewards are designed to recognize, potentially overlooking scientifically valuable candidates that fall outside their definitions.

Defining an appropriate reward is particularly challenging in scientific discovery, where the significance of a material is not always known in advance. Researchers often recognize promising features through their accumulated experience and physical intuition, even when these judgments cannot be expressed as simple numerical criteria~\cite{polanyi1966tacit, kahneman2009conditions}. Whether a compound can be synthesized is judged from experience, and stand-ins such as the distance from the convex hull capture only part of that judgment~\cite{antoniuk2023predicting, jang2020structure}. Features of an electronic structure, such as a flat band, a Dirac cone or a van Hove singularity, are recognized as patterns in a band diagram, and there is no accepted score for them. In machine learning, alignment means making a system pursue what people actually intend, and misalignment is a gap between the objective that a system optimizes and that intent~\cite{leike2018scalable}. In materials discovery, the objective is often a reward written by researchers themselves, such as a score for flat bands, while their intent shows in the judgments they make. Misalignment here is therefore a gap between what researchers can write down and what they actually prefer. A reward can favor candidates that experts would not choose and pass over candidates that they would. In conventional AI applications, reward misalignment is generally regarded as a problem to be corrected, as it can lead systems to pursue unintended outcomes~\cite{amodei2016concrete, skalse2022defining}. Scientific discovery, however, suggests another possibility. What if the disagreement between expert judgment and numerical rewards could itself reveal scientific opportunities that predefined objectives fail to capture? Rather than merely correcting the discrepancy, identifying and exploring it may provide a route toward unexpected materials discovery.

Here, we investigate this possibility using the discovery of electronic flat-band materials as a representative case. 
A flat band near the Fermi level enhances the role of electron interactions, which makes flat-band materials platforms for unconventional superconductivity, magnetism and other correlated phenomena~\cite{checkelsky2024flat}. 
Flatness is therefore a useful indicator for identifying potentially interesting electronic structures, although it does not by itself establish their physical properties. Several numerical descriptors have been proposed to quantify flat bands~\cite{wang2025flatband, regnault2022catalogue}, incorporating features such as bandwidth, proximity to the Fermi level, and the extent of flatness in momentum space. However, these descriptors necessarily impose predefined criteria on what constitutes a desirable flat band. We examine whether such criteria adequately capture expert judgment by comparing a hand-designed bandwidth score with pairwise evaluations collected from researchers in materials science and condensed-matter physics. The researchers choose between two structures rather than scoring each, because people judge more reliably by comparison~\cite{miller1956magical, stewart2005absolute}. Such pairwise choices have been used to learn the intuition of medicinal chemists about molecules~\cite{choung2023extracting}.

\begin{figure}[t]
  \centering
  \includegraphics[width=0.5\textwidth]{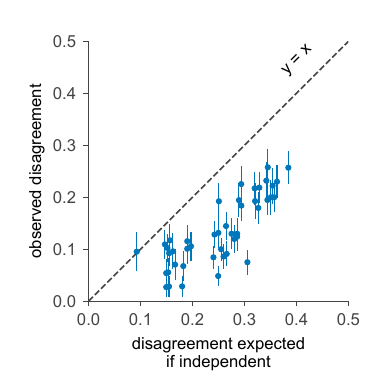}
  \caption{Experts disagree with the bandwidth score in the same direction.
  For each of the $45$ pairs of experts (ten experts judging the
  same $200$ comparisons), the observed disagreement between the two experts
  against the disagreement $e_1(1-e_2) + e_2(1-e_1)$ expected if the two erred
  independently around the score label. Error bars are standard errors, and
  $44$ of $45$ points fall below the diagonal. The count alone does not
  establish a shared criterion. The size of the gap, conditioned on the
  difficulty of the comparisons, does (Supplementary Note~S4).}
  \label{fig:shared-criterion}
\end{figure}

We find that the disagreement between human experts and the numerical bandwidth score is systematic rather than random. In blinded comparisons of electronic band structures, the majority of experts agrees with the score on most pairs, but their disagreements exhibit substantial consistency across individuals (Fig.~\ref{fig:shared-criterion}). %
Although human evaluations inevitably contain mistakes, we statistically estimate that only about $5\%$ of training labels are reversed~\cite{fujisawa2025scalable, fujisawa2026generalised}, insufficient to explain the observed discrepancy. These results suggest that expert judgments contain a shared criterion that is not fully represented by the predefined bandwidth score. We call this systematic difference between what a predefined reward favors and what experts prefer a misalignment.

\begin{figure*}[tbp]
  \centering
  \includegraphics[width=\textwidth]{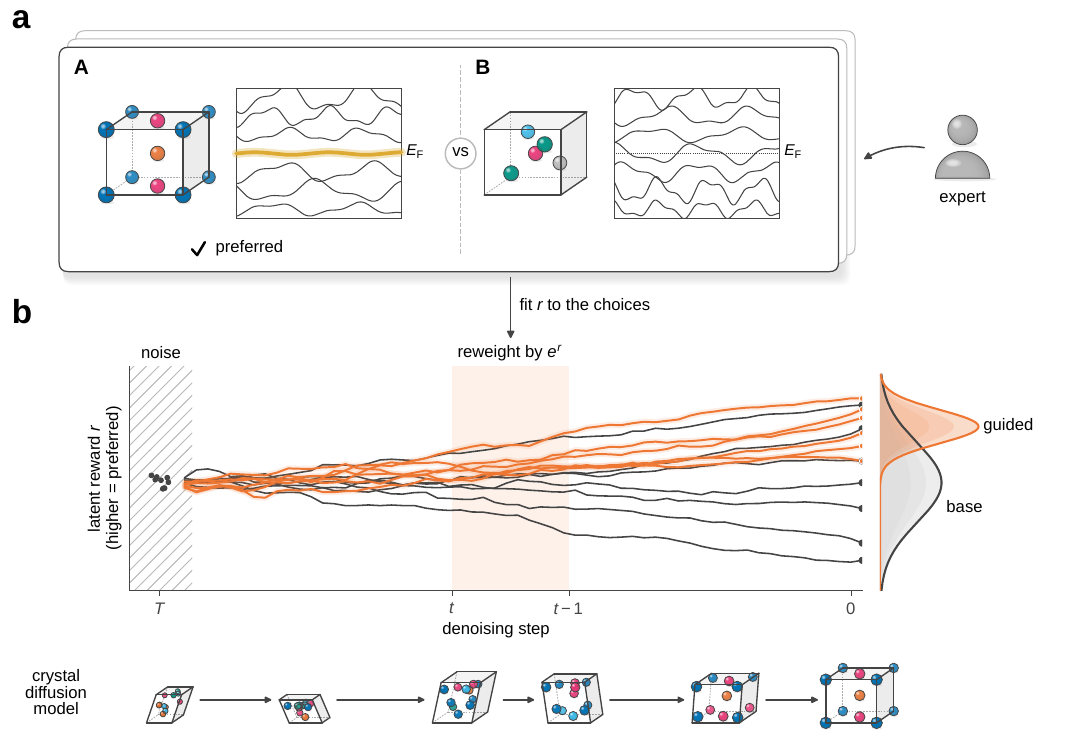}
  \caption{\textbf{Learning expert preferences and guiding generation.}
  \textbf{(a)}~Experts compare pairs of existing crystals from their band structures and record which of the two they prefer. Only the choice within a pair is recorded, so no absolute score is needed. The same procedure accepts preferences from a database record or a computed descriptor (Fig.~\ref{fig:generality}).
  \textbf{(b)}~A latent reward $r$ fitted to the comparisons guides the model, which itself is not retrained. From the same starting noise (hatched), its candidates are reweighted by $e^{r}$ at every denoising step (Methods), so the guided samples (orange) end at higher $r$, the preferred side, than the base samples (grey).}
  \label{fig:motivation}
\end{figure*}

To explore the materials space associated with this misalignment, we employ a preference-learning approach to guide the crystal generator. This approach learns the human latent reward, the unobserved criterion behind the experts' choices, from their pairwise comparisons (Fig.~\ref{fig:motivation}a). Given a preference label indicating that crystal $A$ is preferred over crystal $B$, guidance increases the generation probability of $A$ while decreasing that of $B$ (Fig.~\ref{fig:motivation}b). By learning from these pairwise preferences, the guided generator becomes aligned with human experts' judgments, producing crystals that better reflect their preferences. To make the fit robust to mistaken labels, we combine direct distributional optimization~\cite{kawata2025direct} with a robust H\"older score~\cite{fujisawa2025scalable} and an explicit model of reversed labels~\cite{fujisawa2026generalised}, which discounts the labels the fit cannot explain and estimates their fraction. In blind evaluations, the guided generator produces structures preferred over those of the base model.

Crucially, the learned preferences reveal scientifically promising regions of materials space that the numerical score overlooks. The human latent reward and the bandwidth score largely agree on structures with clearly flat bands, but their rankings diverge for other candidates. Many structures that the latent reward ranks highly are ones that the numerical score does not recognize as flat. A key source of this discrepancy is the rigid threshold imposed by the bandwidth score, which requires a band to remain sufficiently close to the Fermi level over at least half of the sampled momentum path. In contrast, the latent reward also favors structures exhibiting locally flat band segments, which may warrant further investigation even when they fail the conventional criterion. Remarkably, these overlooked structures include previously reported materials such as KCo$_2$As$_2$~\cite{pandey2022, campbell2022} and CoGa~\cite{lau2019spin}, whose electronic structures show a flat band near the Fermi level and a pseudogap, respectively. We also identify new candidate crystals and examine their electronic structures using first-principles calculations. These findings illustrate how exploring the discrepancy between human preferences and numerical rewards can reveal materials candidates that conventional reward-driven screening would discard.

The preference-based framework is also applicable beyond electronic flat bands. We demonstrate its generality using synthesizability preferences derived from experimental records and phonon-related preferences derived from computed descriptors. For synthesizability, a latent reward fitted to the experimental record identifies synthesized structures more often than the distance from the convex hull does. In both cases, pairwise comparisons provide a practical means of learning criteria for materials evaluation and guiding generation. More broadly, our findings suggest a different perspective on reward misalignment in AI-driven scientific discovery. The disagreement between human judgment and numerical rewards is not merely a limitation to overcome; it can also serve as a source of scientific opportunity, revealing promising directions that predefined objectives were not designed to explore.

\begin{figure*}[tbp]
  \centering
  \includegraphics[width=\textwidth]{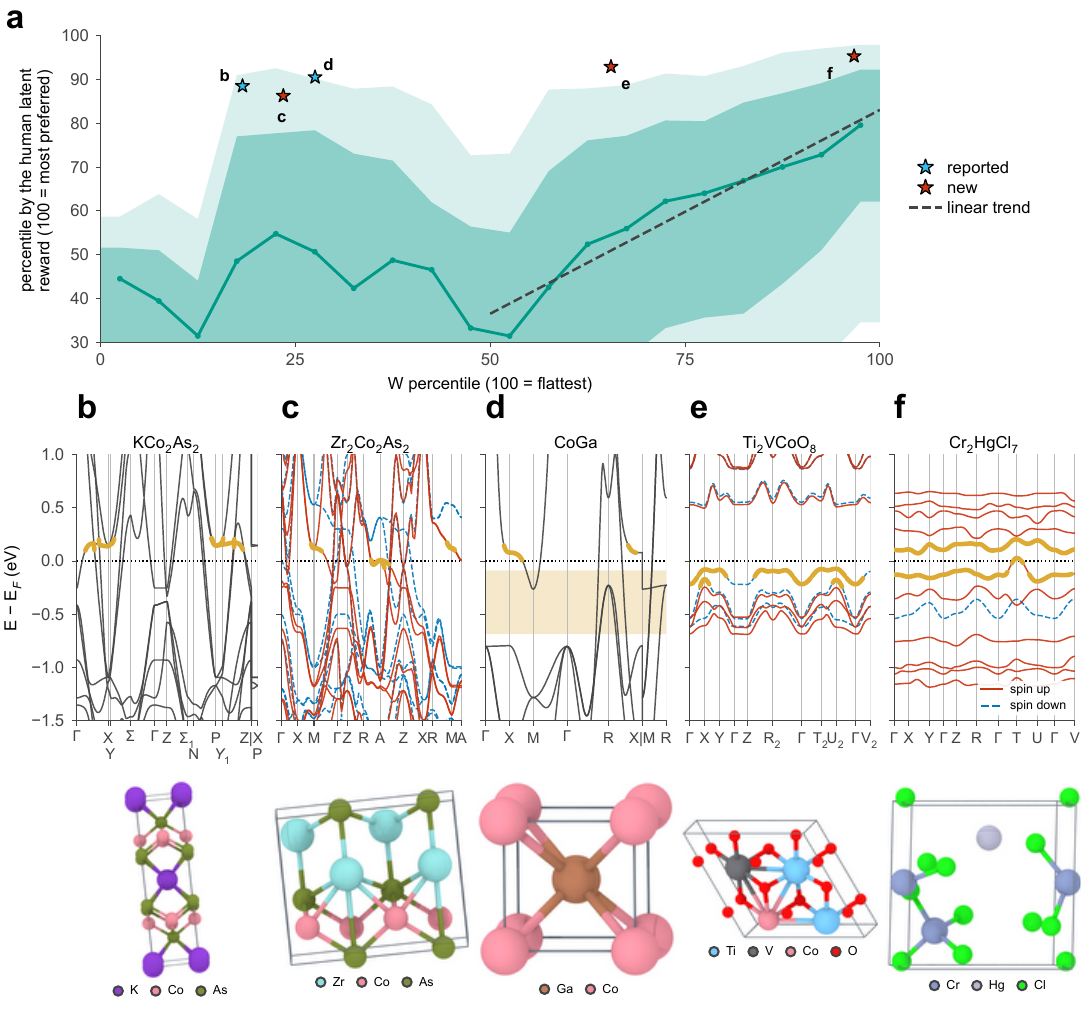}

  \caption{\textbf{The experts' criterion and the bandwidth score agree on the flattest bands and differ below them.}
\textbf{(a)}~Each of the $7{,}958$ relaxed structures from the guided generator ranked by the bandwidth score $W$ and by the human latent reward, as percentiles. The dashed line is a linear fit to the median over the upper half of the $W$ percentiles. The vertical axis starts at $30$ and cuts the lower part of the shaded ranges. Stars, the structures of (b--f) at their ranks, red for the three new candidates and cyan for the two reported materials.
\textbf{(b--f)}~Structure and band structure of the five, from the least to the most flat by $W$. \textbf{(b)}~KCo$_2$As$_2$ ($I4/mmm$) and \textbf{(d)}~CoGa ($Pm\bar{3}m$) are reported materials, with Materials Project PBE bands, spin-degenerate. \textbf{(c)}~Zr$_2$Co$_2$As$_2$ ($P4/nmm$), \textbf{(e)}~Ti$_2$VCoO$_8$ ($Cm$) and \textbf{(f)}~Cr$_2$HgCl$_7$ ($P1$) are new candidates, with CHGNet-relaxed structures and ferromagnetic PBE bands (solid red, spin up; dashed blue, spin down). Shaded in (d), the pseudogap described in the text. In (b--f), thick yellow lines trace each band segment whose mean energy lies within $\pm0.2$~eV of the Fermi level, whose energy varies by at most $0.07$~eV, and which spans at least $4\%$ of the path.}
  \label{fig:mainwr}\label{fig:hddo-candidates}
\end{figure*}

\begin{figure*}[tbp]
  \centering
  \includegraphics[width=\textwidth]{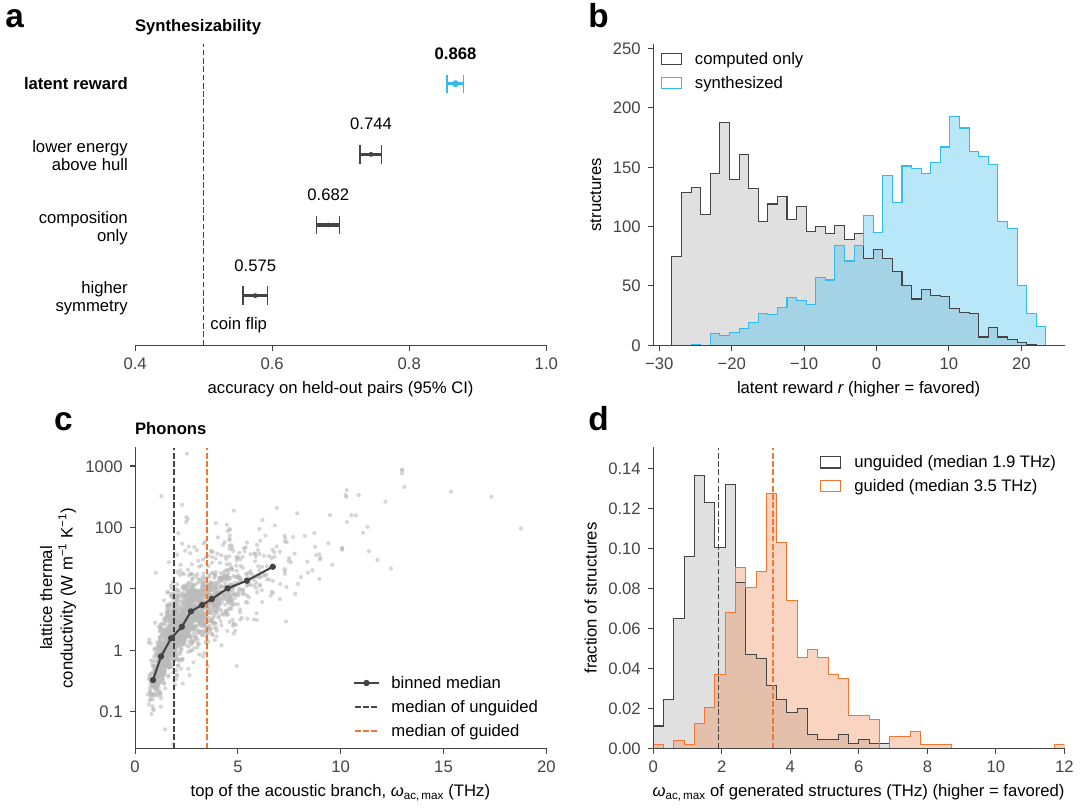}
  \caption{\textbf{Two further case studies with the same procedure.} Top row, synthesizability; bottom row, phonons.
  \textbf{(a)}~Synthesized structures (reported in the Inorganic Crystal Structure Database) are preferred over computed-only structures of the same size. On test pairs whose compositions the latent reward had not seen, the fitted latent reward picks the synthesized one with accuracy $0.868$ $[0.855, 0.879]$ ($n = 2{,}989$ pairs), against $0.744$ for a rule preferring the structure closer to the convex hull, $0.682$ for a composition-only model and $0.575$ for a rule preferring the higher symmetry.
  \textbf{(b)}~The latent reward evaluated on each structure of the same pairs. The two histograms are the two sides of the pairs. This is a descriptive analysis made after the registered evaluation in (a).
  \textbf{(c)}~For the $1{,}954$ training materials with a computed conductivity, the top of the acoustic phonon branch, $\omega_{\mathrm{ac,max}}$, computed from the force constants of the phonon database, against the computed lattice thermal conductivity. The solid line joins the median conductivity in bins of $\omega_{\mathrm{ac,max}}$ holding at least $20$ materials. For $82\%$ $[81, 83]$ of the pairs whose conductivities differ by at least $0.05$ in $\log_{10}$, the one with the higher $\omega_{\mathrm{ac,max}}$ has the higher conductivity. Dashed lines, medians of the generated structures in (d).
  \textbf{(d)}~$\omega_{\mathrm{ac,max}}$ of generated structures, unguided and guided by a latent reward fitted to $\omega_{\mathrm{ac,max}}$ preferences (fraction per bin; $447$ and $486$ valid and scored structures; one guided structure above $12$~THz lies outside the axis). Against unguided draws from the same seeds, guided generation wins $0.746$ $[0.710, 0.779]$ of $600$ comparisons when an invalid structure counts as a loss, and $0.833$ $[0.793, 0.867]$ of the $384$ in which both structures are valid.}
  \label{fig:generality}
\end{figure*}

\section*{Results}

\subsection*{Experts disagree with the bandwidth score in the same direction}
\label{sec:res:criterion}

The first question is whether the experts disagree with the bandwidth score at all, and if so, whether they disagree in the same direction. If each expert's choice were the score's choice plus a random error, the score could replace the experts. To tell random error from systematic disagreement, ten experts judged the same $200$ pairs of band structures (Fig.~\ref{fig:motivation}a), so that their choices can be compared with the score and with one another. Their majority agrees with the bandwidth score $W$ on $83\%$ of the decisive comparisons, those in which neither the majority nor the score is tied. On most comparisons, then, the experts choose as the score does.

The remaining disagreement is not simply random error. If each expert erred independently around the score, any two experts would disagree with each other at a rate fixed by their individual error rates. The observed disagreement falls below that rate for $44$ of the $45$ pairs of experts (Fig.~\ref{fig:shared-criterion}). A shared criterion produces exactly this excess of agreement, and independent errors do not (Supplementary Note~S4).

Mistaken labels cannot explain the disagreement either. Some mistakes are unavoidable in any set of human judgments, from a lapse of attention or from a close call between two similar band structures, so the fitting allows for labels whose choice is reversed and estimates their fraction (Methods). For the $3{,}504$ training labels, that fraction is about $5\%$ (Supplementary Note~S3). Among the votes collected with them, $21\%$ disagree with the score, about four times that rate. The experts thus judge by a criterion of their own, one they can apply but cannot write down. This systematic difference between what the score rewards and what the experts prefer is the misalignment examined in the next subsection.

\subsection*{A generator guided by expert preferences proposes materials the score accepts and materials it rejects}
\label{sec:res:blind}

If the labels carry a criterion, a generator guided by them should produce more of what the experts prefer, and it does. A human latent reward, a function that assigns a value to each crystal, is fitted to the expert comparisons and guides the sampling of the frozen diffusion model (Fig.~\ref{fig:motivation}b, Methods). We report the generator guided by the latent reward fitted to the quality-controlled votes of the training set (Methods). Generators guided or fine-tuned with other choices of labels are reported in Supplementary Note~S7. On pairs of structures generated from the same starting point by the guided and the base model, a blinded rater preferred the guided structure in $65\%$ of comparisons, ties excluded. Judged by the bandwidth score on the same pairs, the guided generator wins $57\%$ (Supplementary Fig.~S3).

The rater and the score agree that the guided generator is the better one, but not on which of its structures are best. Fig.~\ref{fig:mainwr}a ranks the $7{,}958$ structures it generated, after relaxation, both by the bandwidth score and by the human latent reward, with the lowest $W$ and the highest $r$ at the top. Here $W$ is computed from the spin-degenerate bands predicted by the equivariant Hamiltonian network HamGNN~\cite{zhong2023hamgnn}: for each band with at least half of the $k$-points along the path through the Brillouin zone within $\pm 0.2$~eV of $E_F$, we take the difference between its highest and lowest energies at those $k$-points, and $W$ is the smallest such difference ($W = \infty$ if no band qualifies, as for $43\%$ of the structures). The two rankings agree where the bands are clearly flat. Of the $398$ structures in the top $5\%$ by $W$, $77$ are also in the top $5\%$ by the latent reward, against $20$ expected by chance, and the median rank of the $10\%$ flattest structures lies at the $77$th percentile of the latent reward. Below the top, the two rankings diverge, and this divergence is the misalignment seen in the generated structures. The latent reward also ranks highly structures that the score does not count as flat. Of the $796$ structures in its top $10\%$, $221$ have $W = \infty$.

\label{sec:res:candidate}

Fig.~\ref{fig:mainwr}b--f shows five of these structures at the points marked in Fig.~\ref{fig:mainwr}a, two reported materials and three new candidates. KCo$_2$As$_2$
(Fig.~\ref{fig:mainwr}b) has the ThCr$_2$Si$_2$-type structure of the
iron-arsenide superconductors. Its susceptibility rises on warming above $100$~K~\cite{campbell2022, pandey2022}, which has been attributed to a flat band along $\Gamma$--X that produces a sharp peak in the density of states just above $E_F$~\cite{pandey2022}. In the
Materials Project calculation the band lies $0.15$~eV above $E_F$ and the
peak at $+0.26$~eV. CoGa (Fig.~\ref{fig:mainwr}d) is a cubic B2 intermetallic that is paramagnetic in its ordered state~\cite{fecher2022}. Between $-0.69$ and $-0.09$~eV its density of states stays
below half of its maximum between $E_F$ and $+0.3$~eV. This is a pseudogap
whose upper edge lies just below $E_F$, and a band that flattens near X at
$+0.05$ to $+0.10$~eV sits on this edge.

Cr$_2$HgCl$_7$ (Fig.~\ref{fig:mainwr}f) ranks in the top $5\%$ of the $7{,}958$ generated structures both by the bandwidth score ($3.3\%$) and by the latent reward ($4.7\%$). It is ferromagnetic in PBE, with flat majority-spin bands just below and just above $E_F$. The minority channel has no states near $E_F$. Its density of states over the whole Brillouin
zone peaks $0.12$~eV above $E_F$ at $2.2$ times its mean within $\pm 1$~eV. Ti$_2$VCoO$_8$ (Fig.~\ref{fig:mainwr}e) ranks in the top $8\%$ by the latent reward and in the top $35\%$ by $W$. It is a layered oxide whose metal atoms form a single triangular net. In PBE it has
a narrow minority-spin band at $E_F$, and its density of states peaks
$0.03$~eV below $E_F$ at $2.5$ times the mean.

Zr$_2$Co$_2$As$_2$ (Fig.~\ref{fig:mainwr}c) is the candidate on which the two rankings differ most. The latent reward places it in its top $14\%$, but no
band stays within $\pm 0.2$~eV of $E_F$ over half of the path, so
$W = \infty$. Along the path its bands flatten over short segments near $E_F$, at $-0.02$ and $+0.12$~eV. It
adopts the LiFeAs-type arrangement, a cobalt square net tetrahedrally
coordinated by arsenic~\cite{pitcher2008}. In PBE it is a weak ferromagnet
($0.25~\mu_B$ per Co, $2$~meV per ZrCoAs formula unit below the non-magnetic
state). In the non-magnetic state its density of states over the whole Brillouin zone reaches a maximum $0.18$~eV above $E_F$, $1.9$ times $N(E_F)$, close to the flat segments along the path at $+0.12$ to $+0.14$~eV.
Flatness over part of the path
does not by itself guarantee an enhanced density of states at $E_F$, but
combined with the density of states over the whole zone it is a lead for
narrowing the search to materials whose states gather near $E_F$, where
electron correlation is likely to matter~\cite{checkelsky2024flat, stoner1938}.

\subsection*{Two further case studies: synthesizability and phonons}
\label{sec:res:generality}

The procedure applies to any property for which one crystal can be preferred to another, since nothing in it is specific to flat bands. We applied it to two further targets (Fig.~\ref{fig:generality}). The first is synthesizability, whether a proposed crystal can be made in the laboratory. For it we fitted a latent reward and tested whether the latent reward tells synthesized structures from structures that exist only in computation. The second is a property of the phonons, the lattice vibrations that carry heat through a crystal, and for it we fitted a latent reward and used it to guide generation.

A latent reward fitted to the experimental record tells synthesized structures
from computed-only ones better than composition or computed stability does.
A structure was taken as synthesized if it has been reported experimentally
in the Inorganic Crystal Structure Database, and as computed only if it
appears in the Materials Project without such a report. Each pair joined a
synthesized structure with a computed-only structure of the same number of
atoms, and the latent reward was fitted with the same loss as for the expert
labels. On held-out pairs of compositions the latent reward had not seen, it
picks the synthesized structure with accuracy $0.87$. A model that sees only
the composition reaches $0.68$, and a rule preferring the structure computed
to be more stable (closer to the convex hull) reaches $0.74$
(Fig.~\ref{fig:generality}a). The structure itself therefore carries
information about synthesizability beyond what its composition alone
reveals. In a descriptive analysis of the same pairs, scored one structure
at a time, the latent reward also separates the two classes
(Fig.~\ref{fig:generality}b). Of the structures it ranks in the top tenth,
$94\%$ are synthesized, against half of all structures. Pairs of two structures
with the same composition are analyzed in Supplementary Note~S5.

The latent reward matches a published synthesizability classifier. CSLLM~\cite{song2025accurate}, a classifier built on a large language model, was applied to $360$ of the held-out pairs. On these pairs CSLLM reaches an accuracy of $0.84$ and the latent reward $0.86$ (Supplementary Note~S5).

In the phonon case, the criterion was the top of the acoustic phonon branch,
$\omega_{\mathrm{ac,max}}$, with the higher value preferred. Acoustic
phonons carry most of the heat in a non-metallic crystal, and the top of
their branch scales with the speed of sound in the lattice, so crystals with
a higher $\omega_{\mathrm{ac,max}}$ tend to conduct heat
better~\cite{slack1973nonmetallic, toberer2011phonon}. This frequency was
computed from the force constants of a phonon database for the training
corpus, and from a machine-learned model of the interatomic forces for
generated structures. The same descriptor served as the source of
preferences and as the judge of the result. How well the descriptor stands
in for thermal transport can be checked on the training corpus
(Fig.~\ref{fig:generality}c). Among the $1{,}954$ materials of the corpus
with a computed lattice thermal conductivity, in $82\%$ of the pairs whose
conductivities differ clearly, the material with the higher
$\omega_{\mathrm{ac,max}}$ also has the higher conductivity.

A latent reward fitted to pairs ranked by $\omega_{\mathrm{ac,max}}$ guided the
same base model as for flat bands, and the generated structures shift
toward higher values of the descriptor. The median of
$\omega_{\mathrm{ac,max}}$ over the valid and scored generated structures
rises from about $1.9$ to $3.5$~THz (Fig.~\ref{fig:generality}d). Against
base-model structures generated from the same starting points, the guided
generator wins $75\%$ of comparisons when an invalid structure, or one whose
descriptor could not be computed, counts as a loss. Among comparisons in
which both structures are valid, it wins $83\%$. As expected for this
descriptor, guidance favors light, dense compositions (Supplementary
Note~S6). A regression of the computed conductivity on
$\omega_{\mathrm{ac,max}}$ and a few related descriptors of the cell, fitted
to the same corpus, puts the median estimated conductivity of the guided
generator at about $3.7$ times that of the base model.

\section*{Discussion}
We asked whether a hand-designed score captures what researchers look for in a material, and took flat bands as the first case. Experts who compared band structures mostly agreed with the bandwidth score, but their disagreements with the score were systematic rather than random. A human latent reward learned from their comparisons steered a frozen generator toward structures that the experts prefer. It agrees with the score on the flattest structures but diverges below them, and the structures in that region include materials already reported experimentally that the score rejects, and new candidates.

A hand-designed score leaves out part of what experts judge, which is why the two were bound to disagree, and learning from comparisons recovers that part. The experts were not asked to define what makes a band flat, only to say which of two is flatter, and the latent reward fitted to their answers carries whatever criterion lies behind the answers. A criterion, not error, lies behind the answers. The experts agree with one another beyond what independent errors would allow, and mistaken labels are too few to produce the disagreement with the score.

What the experts recognize is a combination of features that no single rule fixes well. How much of a band is flat, how isolated it is from dispersive bands and how close it lies to the Fermi level are each easy to state, and a score must decide in advance how to weigh them against one another. The experts weigh them case by case, and the latent reward learned from their answers without the weighing ever being written down. A hand-designed score keeps only what can be stated as a rule. Comparisons also keep what an expert can recognize but cannot put into words.

What the experts add shows up where the latent reward and the score differ. The latent reward ranks the score's flattest structures highly far more often than chance, so it has learned to recognize clearly flat bands. Below the top the two rankings diverge, and this divergence is the misalignment seen in the generated structures. The guided generator produces both cases, structures that the score also ranks highly and structures that it rejects. Among the structures the score rejects are two materials already reported experimentally that a bandwidth screen would discard, KCo$_2$As$_2$, in the structure of the iron-arsenide superconductors, and CoGa, with a pseudogap just below the Fermi level. The new candidates range from one that both rankings place near the top to one that the score rejects. A generator guided by the experts' preferences thus shows where the experts and the score differ, and which materials lie there. Flatness is an entry point to correlated physics, and the structures that the latent reward favors but the score rejects are where that physics may be found.

The approach is not tied to band structures, since it takes any property for which one crystal can be preferred to another, whether the preference comes from experts, from the experimental record or from a calculation. A latent reward fitted to the experimental record separates synthesized from computed-only structures, and one fitted to a phonon descriptor steers generation toward higher values of that descriptor. Beyond a generator, the approach needs only a judge who can say which of two structures is better, and it does not need that judge to be right every time. Expert preferences are never free of error, and no two experts answer every pair alike, so the fitting is built to recover the shared criterion from answers that are partly wrong.

Learning expert preferences, and what it shows about the misalignment, open several directions. The misalignment itself can serve as a search target. Ranking candidates by how far the learned latent reward and a score disagree would point a search at the structures that experts favor and rules overlook, which is where unexpected physics may appear. The new candidates of Fig.~\ref{fig:mainwr} are the first such targets, and a test in the laboratory is the natural next step. The misalignment also shows what a score would have to capture to come closer to the experts' judgment, for example bands that are flat over only part of the band diagram. The same procedure applies to other features that are recognized by eye, such as Dirac cones and van Hove singularities. As AI systems take over more of the search for materials, comparisons of this kind offer a way to keep expert judgment inside that search.

\section*{Methods}
The procedure, flip-aware H\"older direct distributional optimization (\methodname{}), brings preference learning to crystal generation. It builds on direct preference optimization, which aligns generative models with pairwise human preferences~\cite{rafailov2023direct, wallace2024diffusiondpo}. Following direct distributional optimization~\cite{kawata2025direct}, the aligned distribution is an exponential tilt of the frozen generator, which makes the preference margin exact. A H\"older score makes the fit robust to noisy preferences~\cite{fujisawa2025scalable}, and a generalized Bayesian posterior~\cite{fujisawa2026generalised} with an explicit flip channel estimates the fraction of reversed labels. Unlike preference fine-tuning, the procedure does not retrain the generator. We combine these parts for a crystal diffusion model. We will release the procedure as the open-source software \toolname{} (Code availability).

The procedure has five steps. Pairs of existing crystals are labeled by a source of preference, which can be expert votes, a database record or a computed descriptor (Fig.~\ref{fig:motivation}a). A latent reward $r$, higher for the preferred crystal of a pair, is fitted to the labeled pairs. The difference of its values on the two crystals of a pair is their exact preference margin with respect to the frozen generator, which we call the base model. The fraction of reversed labels, the flip rate, is estimated from those margins. Generation from the base model is then guided by $r$ (Fig.~\ref{fig:motivation}b). Finally, guided and unguided samples are compared in pairs by a blinded rater and by the bandwidth score. The subsections below follow these steps and then describe the first-principles calculations, the two further case studies and reproducibility. Definitions and the full inference equations are in Supplementary Note~S2.

\subsection*{Crystals, pairs, and the bandwidth score}

A crystal $x$ is described by its lattice, the fractional coordinates of its atoms and the atom types. The condition $c$ is its number of atoms, which the generator takes as given. The generator is DiffCSP~\cite{jiao2023diffcsp} in its unconstrained mode, in which the lattice, the coordinates and the atom types are all generated. Each comparison $i$ joins two crystals $x^a_i$ and $x^b_i$ chosen to have the same number of atoms, and so the same condition $c_i$, written in an order that does not depend on the annotation. Its label is $y_i = +1$ if $x^a_i$ was recorded as preferred and $y_i = -1$ otherwise.

Band flatness is scored by the bandwidth score $W$, the bandwidth of the flattest band that stays within $\pm 0.2$~eV of the Fermi level over at least half of the path through the Brillouin zone. The same window is used to build labels and to score outcomes (Supplementary Note~S4). A structure with no qualifying band cannot be scored and receives $W = \infty$. Wherever the score ranks or compares structures, those with $W = \infty$ share the last rank rather than being dropped, because having no flat band is itself a result. Fig.~\ref{fig:mainwr}a spreads these tied structures along its horizontal axis for display, ordered by the energy window that a qualifying band would need, and no value in the text depends on this order.

\subsection*{Preference data}

Two disjoint sets of blind A/B votes on pairs of band structures are used. In the \emph{measurement} set, each of $200$ comparisons was labeled independently by the same ten experts, giving $1{,}596$ decisive votes, that is, votes that are not ties. The experts are researchers with training in materials science or condensed-matter physics. The measurement set is used to characterize the labels. The \emph{training} set has $3{,}504$ decisive votes over $2{,}953$ comparisons. Most of its comparisons carry a single vote, since ten votes per comparison are affordable only for a small set. For $1{,}172$ of these votes, quality control was applied. Experts whose agreement with others fell below $75\%$ over at least $30$ comparisons were excluded. The one expert excluded was never the only one to judge a comparison, so no comparison was lost. The flip rate is estimated on the whole training set, whereas the latent reward that guides the reported generator is fitted to the quality-controlled votes alone. Confidence grades are recorded with every vote and are never used as weights. Folds for cross-fitting (Estimating the flip rate) are formed at the level of comparisons, not votes, so that a comparison with several votes is never split across folds.

\subsection*{Exact terminal-distribution margins}

Let $P_{\mathrm{ref}}(\mathrm{d}x \mid c)$ denote the frozen generator. Its density over final structures is intractable, which is why preference fine-tuning of diffusion models replaces the log-density ratio with a surrogate based on a bound~\cite{wallace2024diffusiondpo}, which can be too loose to control the true objective~\cite{kawata2025direct}. We avoid this substitution by representing the aligned law $Q_r$, the distribution favored by the preferences, as an exponential tilt of the reference by a scalar latent reward $r(x, c)$,
\begin{equation}
  Q_r(\mathrm{d}x \mid c)
  = \frac{\exp\{r(x,c)\}}{Z_r(c)}\, P_{\mathrm{ref}}(\mathrm{d}x \mid c),
  \qquad
  Z_r(c) = \mathbb{E}_{X \sim P_{\mathrm{ref}}}\!\left[e^{r(X,c)}\right].
  \label{eq:tilt}
\end{equation}
The reward network is a copy of the pretrained DiffCSP network with a scalar output head. The latent reward is therefore invariant to permutations of the sites and to rotations of the lattice, and periodic in the fractional coordinates. Because the two structures of a comparison share the condition $c$, the normalizer cancels in their difference, and the difference of their log-density ratios reduces \emph{exactly} to a difference of reward values. That difference is the margin of comparison $i$, and with a learned scale it gives the preference logit,
\begin{equation}
  \Delta_i = r(x^a_i, c_i) - r(x^b_i, c_i),
  \qquad
  z_i(r, b) = e^{b}\, y_i\, \Delta_i,
  \label{eq:exactlogit}
\end{equation}
where $b$ sets the logit scale, which converts the margin into log-odds. The margin contains neither a reference density nor a diffusion lower bound. Without reversals, the probability of the recorded label is $p_i = \sigma(z_i)$, with $\sigma$ the logistic function. The latent reward is thus the unobserved value of a Bradley--Terry model of paired comparisons~\cite{bradley1952rank, thurstone1927law}. The latent reward is fitted by stochastic gradient descent on the loss described in the next subsection. Because the latent reward fits its training comparisons almost perfectly, every reported estimate is made on comparisons that were held out when the latent reward was fitted (cross-fitting; Estimating the flip rate; Supplementary Note~S7). The same fitted latent reward serves twice. Its pairwise differences are the margins from which the flip rate is estimated. As the exponent of the tilt in Eq.~\eqref{eq:tilt}, the latent reward also defines the guided sampler.

\subsection*{Estimating the flip rate}

We model the reversal of a recorded label directly, as a flip channel. Under an independent symmetric reversal of the recorded preference at rate $\epsilon$, the probability of the recorded label is
\begin{equation}
  q_i(r, b, \epsilon)
  = \underbrace{(1-\epsilon)\, p_i}_{\text{kept}}
  + \underbrace{\epsilon\,(1 - p_i)}_{\text{flipped}}
  = \epsilon + (1 - 2\epsilon)\, p_i .
  \label{eq:flipchannel}
\end{equation}
The rate is identifiable only below $\tfrac12$.

The flip rate is inferred in a generalized Bayesian posterior~\cite{bissiri2016general, zhang2006epsilon}, in which a loss takes the place of the negative log-likelihood and a temperature sets how strongly the data weigh against the prior. The loss is the normalized H\"older score of $q_i$~\cite{fujisawa2026generalised} with $\gamma = 0.3$. This score reduces to the cross-entropy as $\gamma \to 0$.

Estimates are made under five-fold cross-fitting. The latent reward is fitted on four folds of comparisons, and the reported flip rate is estimated on held-out comparisons. Scoring in sample understates the rate (Supplementary Note~S7). The latent reward is taken at a fixed training epoch in every fold, since a best-validation checkpoint falls on a different epoch in each fold.

Inference is staged so that the latent reward is fitted only once. The latent reward and the flip rate are first trained jointly on the training comparisons by stochastic gradient descent. The flip rate obtained in this joint fit is a by-product and is not reported. The latent reward is then frozen, and on the held-out comparisons the logit scale is re-estimated together with a provisional flip rate. The temperature is set so that the width of the posterior matches the sampling (sandwich) variance of the corresponding robust estimator. Finally, the posterior in $\epsilon$, marginalized over the logit scale, is evaluated numerically on a grid. It gives the posterior mean and the equal-tailed $95\%$ credible interval (Supplementary Note~S2, Secs.~2.4 and 2.7).

The estimator is checked where the truth is known. We take $2{,}714$ comparisons labeled by the bandwidth score $W$ (median score difference $0.096$~eV). Their structures contain no f-block element, because HamGNN~\cite{zhong2023hamgnn}, the Hamiltonian network used for the band structures (Blind evaluation), does not cover those elements. We then reverse a known fraction $\epsilon_{\mathrm{true}} \in \{0, 0.05, 0.10, 0.20\}$ of the labels at random. The structures are the same in all four conditions, and the estimator sees only the corrupted labels (Supplementary Note~S3).

\subsection*{Guided generation}

The aligned law of Eq.~\eqref{eq:tilt} can be sampled approximately, without changing the generator, by twisted sequential Monte Carlo~\cite{wu2023twisted, zhao2024twisted}. At each reverse diffusion step, $K = 32$ candidate continuations are proposed by the frozen generator and reweighted by a learned twisting function, which estimates how strongly the latent reward will favor each partial structure once it is complete (Supplementary Note~S2, Sec.~2.6). With the reweighting switched off ($K = 1$), the sampler reproduces the frozen generator exactly, structure by structure, on all $199$ structures tested. Each guided structure can therefore be paired with the structure the frozen generator produces from the same initial noise. Because every candidate is drawn from the frozen generator, guidance only reweights what it would produce and cannot create structures it could not.

\subsection*{Blind evaluation}

The guided generator is compared with the base model, the same generator sampled without guidance. For every starting point one comparison is formed from the two structures generated from it, and their band structures, predicted by the equivariant Hamiltonian network HamGNN~\cite{zhong2023hamgnn}, are shown side by side within $\pm 0.4$~eV of the Fermi level. Every comparison was judged by a blinded rater, an author of this study, who did not know which generator produced which structure. The left--right order was randomized at delivery, and a tie option was available. Where a structure from the base model appeared in more than one comparison, those comparisons were not shown close together. Separately, a subset of comparisons was repeated without the rater's knowledge to measure self-consistency. No result was inspected until every vote was in (Supplementary Note~S7). Win rates are computed on decisive votes, with ties excluded. A sensitivity analysis counting each tie as half a win was registered alongside. The bandwidth score judges the same comparisons at its $\pm 0.2$~eV window, narrower than the display window, with unscoreable structures ranked last.

\subsection*{First-principles calculations for the materials of Fig.~\ref{fig:mainwr}}

For the reported materials of Fig.~\ref{fig:mainwr}b,~d we use the band structures and densities of states of the Materials Project~\cite{jain2013materials} (mp-569408 and mp-1121), computed with PBE without spin polarization, with energies referred to the Fermi level of the uniform-mesh calculation (Supplementary Note~S8). Generated structures are relaxed with CHGNet~\cite{deng2023chgnet}, a machine-learned model of the interatomic forces, and their band structures are predicted by HamGNN, as in the evaluation above. For the three new candidates of Fig.~\ref{fig:mainwr}c, e and~f, we instead compute the band structure by density-functional theory with Quantum ESPRESSO~\cite{giannozzi2017qe} and the PBE functional~\cite{perdew1996pbe}, with projector-augmented-wave pseudopotentials~\cite{dalcorso2014pslibrary} and without spin--orbit coupling, on the relaxed structure without further relaxation. Space groups are determined with spglib at a tolerance of $0.1$~\AA{}. Each candidate is computed non-magnetic and ferromagnetic, the latter from a high and a low starting moment. For every candidate both starts converge to the same state, which is shown in Fig.~\ref{fig:mainwr}. %
To test whether the flat segments raise the density of states, we compute it over the whole Brillouin zone from the converged charge density, with the improved tetrahedron method~\cite{kawamura2014tetra} on a $k$ mesh twice as dense as the self-consistent one in each direction. We call the density of states enhanced at $E_F$ if, after Gaussian smoothing with a width of $0.02$~eV, it has a local maximum within $\pm 0.15$~eV of $E_F$ that exceeds $1.5$ times its mean within $\pm 1$~eV. This rule was fixed before the calculations. Settings, convergence checks and the densities of states are given in Supplementary Note~S8.

\subsection*{Synthesizability case study}

The synthesizability preference labels are derived from database records. The entries come from the MP-20 set~\cite{xie2022cdvae}, the subset of the Materials Project on which the generator was trained. Only entries with 4--20 atoms in the primitive cell and a clear record of whether they were reported experimentally are used. Of two entries with the same number of atoms in the primitive cell, the one reported experimentally in the Inorganic Crystal Structure Database (synthesized) is preferred to the one without such a report (computed only). Matching the cell size removes a trivial cue. There are $40{,}839$ usable entries, $18{,}370$ synthesized and $22{,}469$ computed only. Their compositions are split 80/20 before pairing, so that no composition is on both sides. The 80\% part gives $12{,}248$ training pairs and the 20\% part gives $2{,}989$ held-out pairs of unseen compositions. A second held-out set, of $690$ pairs of polymorphs (different structures of one composition), comes from compositions withheld from training entirely.

As in the flat-band case, the latent reward is the DiffCSP network with a scalar output. It is trained on the training pairs with the same H\"older loss for $100$ epochs (seed 42, with two further seeds as a check). It is compared with a logistic regression on the element fractions of the composition and with two rules, which prefer the structure with the lower computed energy above the convex hull, a measure of stability, or the one with the higher space-group symmetry. The decision rule was fixed before scoring, with two tests. The first compares the latent reward with the composition-only regression on pairs of new compositions, and the second compares the latent reward with the hull rule on pairs of the same composition. In each test the latent reward is more accurate if the $95\%$ interval of the paired difference in accuracy lies above zero, less accurate if it lies below zero, and otherwise not distinguishable. The latent reward only scores held-out structures and guides no generation (Supplementary Note~S5).

\subsection*{Phonon case study}

The phonon preference is defined by a computed descriptor, $\omega_{\mathrm{ac,max}}$, the top of the acoustic phonon branches, and of two structures the one with the larger value is preferred. The corpus for the latent reward for phonons is built from the MDR phonon calculation database~\cite{togo2025mdrphonon}, which supplies harmonic force constants, and the Phonix database~\cite{ohnishi2026phonix}, which supplies lattice thermal conductivities. The corpus holds the $1{,}973$ materials of 4--20 atoms with converged harmonic phonons without imaginary modes, of which $1{,}954$ also have a computed conductivity (Fig.~\ref{fig:generality}c).
The value of $\omega_{\mathrm{ac,max}}$ for each material is computed with phonopy~\cite{togo2023phonopy} from the force constants in the database. Pairs are formed only between materials with the same number of atoms, and pairs whose descriptors differ by less than $0.1$ standard deviations are dropped. The materials are split into five folds before pairing, so that no material appears in two folds, giving $31{,}693$ pairs. Four folds train the latent reward for phonons with the loss of the flat-band case, for $100$ epochs. The fifth fold is held out as a check. The run would have been stopped had the latent reward misordered more than $20\%$ of the pairs in this fold, and the latent reward misordered $5.4\%$. Guided and base-model structures are drawn in $600$ pairs over three seeds, each pair generated from the same starting point, with the sampler of the flat-band case (settings that differ are in Supplementary Note~S6). Each structure is scored with phonopy, using force constants from the machine-learned model of the interatomic forces MatterSim~\cite{yang2024mattersim} and the descriptor settings of the corpus. A guided structure wins a pair if its $\omega_{\mathrm{ac,max}}$ is larger. Two win rates are reported. The primary one, chosen by a rule registered before scoring, counts an invalid structure, or one whose descriptor could not be computed, as a loss. The second uses only pairs in which both structures are valid and scored. Validity criteria and the regression that estimates conductivity from the descriptors are in Supplementary Note~S6.

\subsection*{Reproducibility}

Generation is deterministic: rerunning a job with the same seed and settings reproduces the same structures exactly, on every node and GPU type tested (Supplementary Note~S7).

\section*{Code availability}

\toolname{}, which implements \methodname{}, will be made publicly available at
\url{https://github.com/RyotaroOKabe/CrysTacit} under the MIT license. The generator is adapted from DiffCSP~\cite{jiao2023diffcsp}.

\section*{Acknowledgement}
This work was supported by JST, ARiSE, Japan Grant Number JPMJAR2628. This research used resources of the National Energy Research Scientific Computing Center (NERSC), a Department of Energy User Facility using NERSC award DDR-ERCAP0037478.

\clearpage

\bibliography{main}

\includepdf[pages=-]{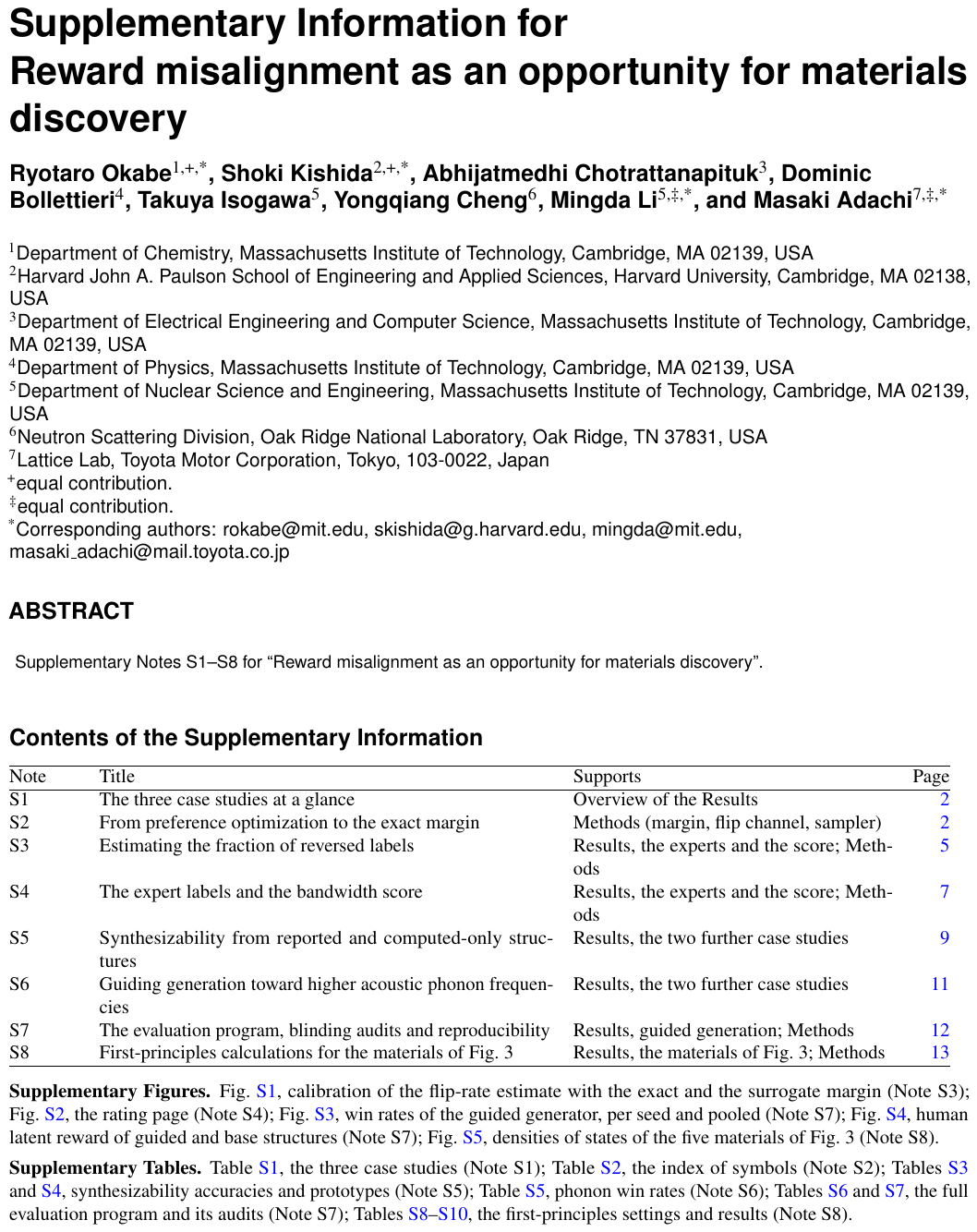}
\end{document}